\documentclass[conference, final]{IEEEtran}
\IEEEoverridecommandlockouts
\usepackage{amsmath,amssymb,amsfonts}
\usepackage{algorithmic}
\usepackage{graphicx}
\usepackage{textcomp}
\usepackage{orcidlink}
\usepackage{bm}
\usepackage{subfigure}
\usepackage{hyperref}
\newcommand{\angstrom}{\mbox{\normalfont\AA}}

\def\BibTeX{{\rm B\kern-.05em{\sc i\kern-.025em b}\kern-.08em
    T\kern-.1667em\lower.7ex\hbox{E}\kern-.125emX}}
\begin{document}

\title{Multi-spectral Entropy Constrained Neural Compression of Solar Imagery}

\author{
    Ali Zafari$^\dag$
    , Atefeh Khoshkhahtinat$^\dag$, Piyush M. Mehta$^\ddag$, Nasser M. Nasrabadi$^\dag$\\
    Barbara J. Thompson$^\S$,
    Michael S. F. Kirk$^\S$, Daniel da Silva$^\S$\\
    $^\dag$Department of Computer Science \& Electrical Engineering, West Virginia University, WV USA\\
    $^\ddag$Department of Mechanical \& Aerospace Engineering, West Virginia University, WV USA\\
    $^\S$NASA Goddard Space Flight Center, MD USA\\
    {
    \tt\small \{\href{mailto:az00004@mix.wvu.edu}{az00004},\quad\href{mailto:ak00043@mix.wvu.edu}{ak00043}\}@mix.wvu.edu,\quad\{\href{mailto:piyush.mehta@mail.wvu.edu}{piyush.mehta},\quad\href{mailto:nasser.nasrabadi@mail.wvu.edu}{nasser.nasrabadi}\}@mail.wvu.edu
    }\\
    {
    \tt\small \{\href{mailto:barbara.j.thompson@nasa.gov}{barbara.j.thompson},\quad\href{mailto:michael.s.kirk@nasa.gov}{michael.s.kirk},\quad\href{mailto:daniel.e.dasilva@nasa.gov}{daniel.e.dasilva}\}@nasa.gov
    }
}


\maketitle

\begin{abstract}
Missions studying the dynamic behaviour of the Sun are defined to capture multi-spectral images of the sun and transmit them to the ground station in a daily basis. To make transmission efficient and feasible, image compression systems need to be exploited. Recently successful end-to-end optimized neural network-based image compression systems have shown great potential to be used in an ad-hoc manner. In this work we have proposed a transformer-based multi-spectral neural image compressor to efficiently capture redundancies both intra/inter-wavelength. To unleash the locality of window-based self attention mechanism, we propose an inter-window aggregated token multi head self attention. Additionally to make the neural compressor autoencoder shift invariant, a randomly shifted window attention mechanism is used which makes the transformer blocks insensitive to translations in their input domain. We demonstrate that the proposed approach not only outperforms the conventional compression algorithms but also it is able to better decorrelates images along the multiple wavelengths compared to single spectral compression.
\end{abstract}


\section{\textbf{Introduction}}
\begin{figure*}[tp]
    \centering
    \includegraphics[width=0.95\linewidth]{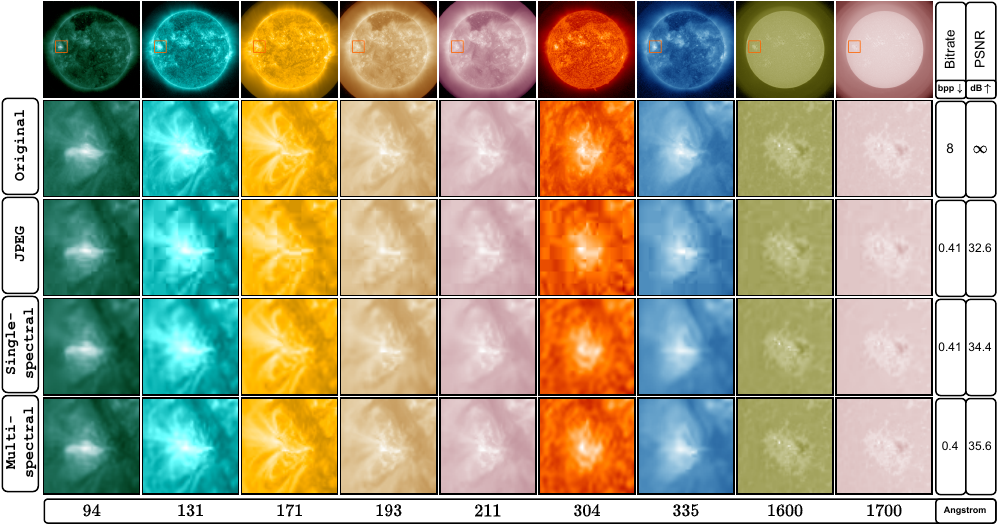}
    \caption{Visual comparison of compressing multi-spectral images using our proposed multi-spectral compression against traditional image compression JPEG and also a neural compression which compress images single-spectrallly. Proposed multi-spectral compression can achieve the best rate-distortion performance [bpp$\downarrow$/PSNR$\uparrow$]. Section \ref{sec:methods} discusses more about the advantages of multi-spectral compression over other methods. \emph{Best viewed on screen.}}
    \label{fig:visual-comparison}
\end{figure*}
Data compression is inevitable when there is a need for transmitting huge amount of data in a limited bandwidth. Neural network based compression algorithms have shown great potential on replacing traditional codecs \cite{yang2023introduction}. Instead of relying on hand-engineered linear transforms, \emph{e.g.}, DCT in JPEG \cite{wallace1992jpeg}, neural compressors can be trained on an arbitrary set of images depending on the task in hand. This major advantage encourages their usage in any other field to reach better trade-off for the ad-hoc application of data compression algorithms.

Data intensiveness of space missions studying the Sun for the goal of space weather analysis and prediction, sets stringent constraints on the bandwidth usage for data communication \cite{chamberlin2012sdobook}. To meet the bandwidth requirements in such missions, lossy compression should be investigated which opens a lot of room for efficient data transmission.
As an example, Solar Dynamics Observatory (SDO) captures images of the Sun at the resolution of $4096\times4096$ in nine different wavelengths at a cadence of 12 seconds which results in transmitting more than 1.4 terabytes of data each day to the the ground station \cite{lemen2012aia}. Transmitting this huge amount of data justifies the investigation for a multi-spectral compression algorithm to remove redundancies over the different frequencies and effectively save transmission cost \cite{zafari2022attention, zafari2022neural}.

Convolution-based neural compression systems \cite{balle2016, minnen2020} although showing great potential in replacing traditional codecs, are being replaced by transformer-based networks \cite{bai2022towards,zhu2022trasformbased,zou2022window,liu2023mixed} after the successful application of vision transformers in the computer vision community \cite{dosovitskiy2021vit,liu2021swin}. Despite their great ability to capture long-range dependencies, to have a feasible implementation, their global self-attention mechanism must be bridled which comes with a deterioration in performance. This performance degradation need to be addressed properly to make use of these powerful architectures in the low-level computer vision algorithms \cite{xia2022deformablevit,tu2022maxvit, li2022uniformer, zhang2022crossformer}.

To get the most out of a multi-spectral neural image compressor we used transformers with window-based local self attention which are modified to achieve better rate-distortion performance.
First by aggregating the keys and queries inter-windows we unleash  limited capacity of local window attention mechanism to let the transformer module decide on which token outside its current window to attend. By doing so, we have kept the computation complexity not increased quadratically and simultaneously, enhanced the global dependency capturing of the model.
One other important direction is to make the transformer network be able to be shift invariant with respect to its input. Shifted window self-attention \cite{liu2021swin,hatamizadeh2023globalcontext, wang2022uformer} is not able to preserve this vital feature for the task of image compression. To make the transformer network insensitive to shifted input, we sample the shift size randomly during training which enforces the network not to distinguish between different shift sizes and operate in a translation invariant mode.

As a sample visualization of our multi-spectral data and the achieved rate-distortion performance of the proposed method, we refer to Figure \ref{fig:visual-comparison}. For further discussions and analysis, this paper is organized as follows: 
Section \ref{sec:related-work} provides a review of neural compression autoencoders and their potential application for a solar mission, with a specific focus on SDO mission and its multi-spectral data.
Section \ref{sec:methods} presents our proposed method, highlighting the architecture, aggregated window self attention mechanism, and shift invariance transformer blocks.
Experimental results and ablation studies are described in Section \ref{sec:experiments}, showcasing the performance of our method on the SDO dataset.
Section \ref{sec:conclusion} concludes the paper by summarizing our findings.

\section{\textbf{Related Work}}\label{sec:related-work}
In the first part of this Section we will review the works done in the field of neural image compression and the usage of transformers, then in the second part we focus on the solar dynamics observatory mission.
\subsection{\textbf{Rate-Distortion Optimized Neural Compression}}
\textbf{Rate-Distortion Variational Autoencoders} (RD-VAE) have dominated the neural-based learned image compression algorithms \cite{theis2017, balle2016, balle2017endtoend, balle2018a, yang2023introduction}. The idea is to replace the Gaussian prior and posterior with uniform distribution to have the hard scalar quantization be simulated by adding unit uniform noise to the bottleneck \cite{balle2018b, balle2021}. Another modification to the vanilla VAEs will be to add a learnable prior entropy model \cite{balle2018a} and use it to measure the cross entropy between the true distribution of the latent code and the shared prior between encoder and decoder. We follow the same paradigm but with transformers as nonlinear analysis and synthesis transforms \cite{zou2022window, zhu2022trasformbased, bai2022towards, zafari2023frequency}.

\begin{figure*}[tp]
    \centering
    \includegraphics[width=0.9\linewidth]{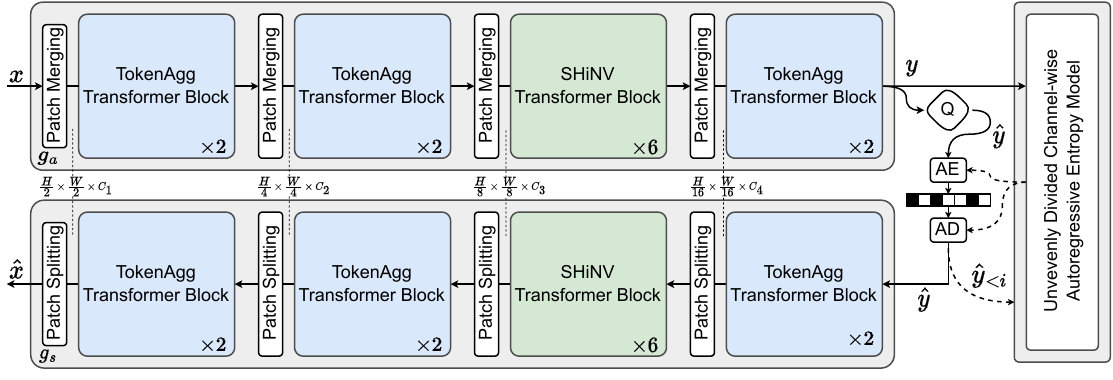}
    \caption{Transformer-based network architecture. Multi-spectral image $\bm x$ is fed into consecutive transformer blocks to generate the scalar quantized bottleneck ($\bm{\hat{y}}$). Token dimensions in each stage are defined as $(C_1, C_2, C_3, C_4)=(160, 256, 352, 448)$.
    Afterwards the quantized latent is further compressed/decompressed losslessly by arthmetic encoder/decoder (AE/AD), respectively. The black and white checkerboard box represents the bitstream of compressed latent features. Decoded latent is then used to reconstruct the lossy compressed image ($\bm{\hat{x}}$). To estimate the entropy of the latent code, a channel-wise autoregressive model is utilized which is divided unevenly over the channel dimension in which each division of latent code $\bm{\hat{y}_{i}}$ has a probability dependent on its previous divisions $\bm{\hat{y}_{<i}}$. Analysis and synthesis transformer-based nonlinear transforms are denoted by $g_a$ and $g_s$, respectively. Token Aggregated (TokenAgg) and SHift iNVariance (SHiNV) transformer blocks are described in Section \ref{sub:transformers}.}
    \label{fig:architecture}
\end{figure*}
\textbf{Vision Transformers} \cite{dosovitskiy2021vit} have shown a great potential in replacing the CNN-based transforms in the learned transform coding architectures. Authors in \cite{bai2022towards} applied transformer blocks only in the main analysis and synthesis transforms. \cite{zou2022window} proposed to use the idea of local window self-attention to improve the performance of the CNN architecture. Augmenting both main and hyper transforms has shown better rate-distortion trade-off \cite{zhu2022trasformbased} by using local self attention augmented by shifting windows as first proposed under the name of Swin transformer block \cite{liu2021swin}. Sequential \cite{lu2022dcc} and Parallel-wise \cite{liu2023mixed} mixing of transformer blocks with CNNs were studied as well, where the latter presented the state of the art in terms of rate-distortion performance of general neural image compression. 

The Transformer model possesses several advantageous properties for constructing robust data-driven models. Firstly, it can effectively capture dependencies that span long distances within the data \cite{vaswani2017transformer}. Secondly, it exhibits minimal inductive bias, enabling greater flexibility in accommodating vast amounts of data \cite{dosovitskiy2021vit}. Lastly, its high parallelism greatly benefits the training and inference processes of large-scale models \cite{devlin2018bert,radford2018gpt,vaswani2017transformer}. Consequently, the Transformer has not only revolutionized natural language processing but has also displayed promising advancements in the field of computer vision.

Within the computer vision community, there has been a significant proliferation of vision transformers recently \cite{liu2021swin, tu2022maxvit, wang2021pvit} . Notably, self-attention, the fundamental building block of these models, has been a popular subject of research \cite{choromanski2021performer,zhu2023biformer,taghaviliquid,xiao2022stoformer, schober2023stochastic}. Unlike convolution, which is inherently localized in its operations, attention's pivotal characteristic is its global receptive field \cite{ramachandran2019stand, choromanski2021performer, rahman2023multivariate, nematirad2023forecasting, li2022uniformer}. This empowers vision transformers to capture long-range dependencies \cite{cordonnier2020relation}. However, this advantageous feature comes at the expense of increased computational complexity and memory requirements, as attention calculates pairwise token affinity across all spatial locations, resulting in substantial memory footprints. This work proposes two enhancements upon local shifted window vision transformers, as thoroughly explained in Section \ref{sub:transformers}.

\subsection{\textbf{Space Missions Studying the Solar Atmosphere}}

The advancement of sensor technology and the growing need for a deeper understanding of the space environment, ranging from the Sun to Earth and beyond, has resulted in a massive increase in data volume. This includes data with unprecedented spatial and/or temporal resolution, as well as multi-spectral information. Consequently, innovative data compression algorithms are required to handle this vast amount of data efficiently.
Solar Dynamics Observatory (SDO) is a recently designated space mission to study the atmosphere of the Sun by observing it continuously. SDO carries Atmospheric Imaging Assembly on-board which captures 4K resolution images at 10 different wavelengths at a casdence of 10 seconds \cite{lemen2012aia}. Downloaded imges includes seven Extreme UltraViolet (EUV) bands of $94, 131, 171, 193, 211, 304, 335\,\angstrom$ in addition to two visible wavelengths $1600$ and $1700 \,\angstrom$ \cite{sdoguide}. This instrument merely transmits about 1.4 Tera bytes of data each day orbiting the earth \cite{chamberlin2012solar,lemen2012aia}. This huge amount of imagery data calls for compression mechanisms to be introduced specially designed for these types of data-intensive missions \cite{sdoguide, sarlak2021approach, bhuveladesign}.
The crucial need for lossy image compression on petabyte-scale data obtained from solar missions has been emphasized by applying JPEG-2000 \cite{fischer2017jpeg2000eve} and learning-based \cite{zafari2022attention,zafari2022neural, zafari2023ams} methods to compress SDO images although non has considered utilizing the redundancies over the spectrum dimension, which is the topic of this work.

\textbf{Learning-based analysis of solar data} has gained attention recently by the work of Galvez et al. \cite{galvez2019}, which provided a set of SDO data ready for machine/deep learning analysis. A portion of raw data from the Solar Dynamics Observatory (SDO) was collected and processed to create a machine-learning ready dataset known as SDOML. This dataset was specifically curated to facilitate the development and evaluation of learning-based methods using SDO mission data.

Building upon the SDOML dataset, Salvatelli et al. \cite{salvatelli2022} employed a U-Net architecture within a Generative Adversarial Network (GAN) framework. Their approach aimed to translate multi-spectral images from the Atmospheric Imaging Assembly (AIA) instrument of SDO, captured at wavelengths 94, 171, and 193 \angstrom, to a specific target wavelength of 211 \angstrom. This work focused on spectral image translation using deep learning techniques.
Another machine learning study conducted on the SDOML dataset was proposed by Santos et al. \cite{santos2020}. They utilized deep neural networks to address the auto-calibration of instrument degradation in SDO imagery. By leveraging the SDOML dataset, their approach aimed to automatically compensate for degradation effects in the instrument imagery, improving the accuracy and reliability of the captured data.
In a different application, Dash et al. \cite{dash2021super} employed a conditional GAN to perform image translation from the Helioseismic and Magnetic Imager (HMI) images downloaded from SDO to AIA images. This translation allowed them to generate AIA-like images from the HMI instrument, thus expanding the capabilities of HMI images through the use of deep learning techniques.


\section{\textbf{Methods}}\label{sec:methods}

\subsection{\textbf{Tansform Coding Neural Image Compression}}
Autoencoder-based image compression networks \cite{balle2018a, minnen2018}, such as the illustrated architecture in Figure \ref{fig:architecture}, typically comprise two main components. The first component is the encoder/decoder network, while the second component is the bottleneck entropy modeling network. The details of the latter network are extensively explained in Section \ref{section:entropy-model}. Referring to Figure \ref{fig:architecture}, we can summarize the relationship between the network input ($\bm{x}$) and output ($\bm{x'}$) as follows:
\begin{equation}
 \begin{aligned}
    &\bm{x'}=g_s(\bm{\hat{y}};\bm{\theta_g}),\\
    &\bm{\hat{y}}=\lfloor g_a(\bm{x};\bm{\phi_g})\rceil,\\
    &\bm{\hat{z}}=\lfloor h_a(\bm{y};\bm{\phi_h})\rceil,
\end{aligned}
\end{equation}
where the output image $\bm{x'}$ is generated by applying the synthesis transform $g_s$ to the quantized latent variable $\bm{\hat{y}}$, controlled by the learned parameters $\bm{\theta_g}$. The quantized latent variable $\bm{\hat{y}}$ is obtained by quantizing the output of the analysis transform $g_a$ applied to the input image $\bm{x}$ using the learned parameters $\bm{\phi_g}$. The quantized hyper-prior $\bm{\hat{z}}$ is obtained by quantizing the output of the analysis transform $h_a$ applied to the latent variable $\bm{y}$, using the learned parameters $\bm{\phi_h}$. The subscripts $a$ and $s$ indicate that $g_a$ and $g_s$ represent analysis and synthesis transforms, respectively, commonly used in the terminology of transform coding-based compression algorithms.

\subsubsection{\textbf{Entropy-Constrained Distortion Minimization}} \label{sec:compression:objective}
Any learned image compression network aims to balance rate and distortion, which is represented by the Lagrangian paramtere $\lambda$ in the equation:
\begin{equation}
    R+\lambda D
    \label{eq:rate-distortion},
\end{equation}
where $R$ represents the estimated entropy of the latent code, and $D$ corresponds to the reconstruction distortion. During network training, the goal is to minimize the rate term, which is the estimated entropy of the quantized bottleneck. The probability distribution of the latent code is approximated by the hyper-prior $\bm{z}$. The quantized $\bm{\hat{z}}$ is transmitted along with the compressed image as side-information. Thus, both the entropy of the latent code and the hyper-prior need to be optimized, as defined below:
\begin{equation}
    R=\mathbb{E}_{x\sim p_X}[-\log_2P_{\bm{\hat{y}}|\bm{\hat{z}}}(\bm{\hat{y}}|\bm{\hat{z}};\bm{\theta_h})-\log_2P_{\bm{\hat{z}}}(\bm{\hat{z}};\bm{\psi})],
\end{equation}
In Equation \ref{eq:rate-distortion}, $D$ represents the distortion between the input and output images of the network, which can be measured using various metrics. The Mean Squared Error (MSE) is commonly used but has been criticized for producing blurry reconstructions. Alternative metrics such as Multi Scale Structural Similarity Index (MS-SSIM) have been proposed to align with the human visual system but have their limitations when closely examined.


\subsubsection{\textbf{Bottleneck Entropy Modeling}} \label{section:entropy-model}
The performance of a learned image compression scheme relies heavily on its ability to accurately estimate the true entropy of the bottleneck. The objective is to minimize the cross-entropy between the estimated and true entropies. Various probability estimation methods have been proposed in the literature, including empirical histogram density estimation \cite{theis2017}, piecewise linear models \cite{balle2017endtoend}, conditioning on a latent variable (hyper-prior) \cite{balle2018a}, and context modeling based on autoregressive models \cite{minnen2018}.

At a high level, entropy estimation models can be categorized into two main types: Forward Adaptation (FA) models and Backward Adaptation (BA) models. FA models have limited capacity to capture all dependencies in the probability distribution of the latent code, while BA models suffer from the inability to parallelize the decoding process. Learned FA models utilize only the information provided during the encoding of the image, whereas BA methods based on autoregressive models require information from the decoded message as well. To leverage the advantages of both types of models, a conditional probability formulation is defined:
\begin{equation}
P_{\bm{\hat{y}}|\bm{\hat{z}}}(\bm{\hat{y}}|\bm{\hat{z}})=\prod_i P(\bm{\hat{y}_i}|\bm{\hat{y}_{j<i}},\bm{\hat{z}};\bm{\theta_h}).
\end{equation}
Conditioning on the quantized hyper-prior $\bm{\hat{z}}$ as side-information represents a form of FA, while conditioning on all previously decoded elements of the latent space $\bm{\hat{y}_{j<i}}$ represents a form of BA.

The performance of BA models has been improved in \cite{minnen2020} by introducing conditioning only between slices of channels in the bottleneck. Unlike spatial autoregressive modeling in \cite{minnen2018}, \cite{minnen2020} considers conditioning probabilities only on the channels. This approach enables reasonable parallelization of the decoding process. We have employed the same approach as \cite{minnen2020}, but by dividing channels in an uneven set of groups \cite{he2022elic}, to estimate and minimize the entropy during training as shown in Figure \ref{fig:architecture}.

\subsection{\textbf{Transformer-based Nonlinear Transforms}}
\label{sub:transformers}
Nonlinear Transform coding \cite{balle2021} will be the choice when it comes to complex data distributions such as natural images or, in our case, multi-spectral images. Transformers \cite{vaswani2017transformer, dosovitskiy2021vit}, which are based on self-attention mechanism, are replacing traditional convolution-based backbones in the deep neural networks used in computer vision tasks \cite{mohamadi2022deep,liu2021swin, tu2022maxvit}.
Although used as powerful representation learner, two issues of the self-attention need to be addressed to perform well when it comes to the task of image compression. First its notorious quadratic computation complexity which which is addressed in the literature by sparsifying global attention, limiting it to a local window \cite{liu2021swin}, widening the window by dilation \cite{zhang2022crossformer} and make the structure hierarchical \cite{wang2021pvit}. Second issue which is more concerned for the task of image compression is the translation invariance. In the task of compression, in contrast to object detection or classification, it is necessary for the encoder and decoder nonlinear transforms to extract representative features of the whole image no matter of its spatial location. Even in contrast to object detection, minute local details could be of more importance compared to low frequency predictable regions. Here we briefly set the terminology of self-attention mechanism and in sections \ref{subsub:aggregated} and \ref{subsub:shinv} describe our proposed solutions to ameliorate its performance.


\subsubsection{\textbf{(Preliminaries) Self-Attention Mechanism}}
\label{subsub:self-attention}

Let's assume we have an input $\mathbf{X}$ of size $\mathbb{R}^{N\times C}$. To apply the self-attention machanism on $N$ tokens of dimension $C$, three linear transformations of it are calculated which are called query ($\mathbf{Q} \in \mathbb{R}^{N\times C}$), key and value ($\mathbf{K},\mathbf{V} \in \mathbb{R}^{N_{KV}\times C}$). Then the self-attented input will be as follows:
\begin{align}
    \text{SA}(X) = \text{Softmax}(\frac{\mathbf{QK}^T}{\sqrt{C}})\mathbf{V}.
\end{align}
More commonly, the self-attention is divided into multiple heads to bring different aspects of the input into attention. Having each head $\mathbf{H}_i \in \mathbb{R}^{N\times \frac{C}{n}}$, its corresponding query, key and value are calculated as
\begin{align}
\begin{split}
    &\mathbf{Q}_i = \mathbf{Q}\mathbf{W}^Q_i,\\
    &\mathbf{K}_i = \mathbf{K}\mathbf{W}^K_i,\\
    &\mathbf{V}_i = \mathbf{V}\mathbf{W}^V_i,
\end{split}
\end{align} 
and then each head is self-attended as
\begin{align}
    \mathbf{H}_i=\text{Softmax}(\frac{\mathbf{Q}_i\mathbf{K}_i^T}{\sqrt{C/n}})\mathbf{V}_i,
\end{align}
where linear weights for each head are $\mathbf{W}^Q_{1\dots n},\mathbf{W}^K_{1\dots n},\mathbf{W}^V_{1\dots n} \in \mathbb{R}^{C\times \frac{C}{n}}$. To have the final output of multi-head self-attention, all the heads need to be concatenated and passed through a final linear ($\mathbf{W^H} \in \mathbb{R}^{C\times C}$) transformation as follows:
\begin{align}
    \text{MHSA}(\mathbf{X})=[\mathbf{H}_1;\dots;\mathbf{H}_{n}]\mathbf{W}^H.
\end{align}
To keep the equations uncluttered, we resume our discussion by only having a single head, while in practice our work can be easily extended to multi-heads.\\

\subsubsection{\textbf{Inter-Window Token Aggregated}}
\label{subsub:aggregated}

\begin{figure}[tp]
    \centering
    \includegraphics[width=0.8\linewidth]{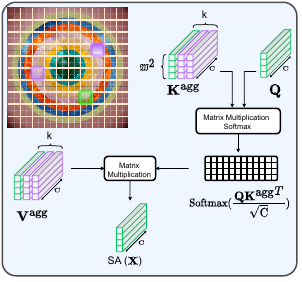}
    \caption{Inter-window token aggregated attention module. Top-k closet windows (assumed $k=3$, $\mathfrak{W}=2$ for visualization) ({\color{violet}{violet}}) to be attended for the the targeted window ({\color{green}{green}}). (\emph{multi-spectral input is shown as hyper-disks for visualization purposes only.})}
    \label{fig:inter-window}
\end{figure}

Window-based self-attention \cite{liu2021swin} with a fixed window size of $\mathfrak{W}$ first partitions tokens into groups of size $\mathfrak{W}^2$ resulting in $\mathbf{X}^\mathfrak{W} \in \mathbb{R}^{\frac{HW}{\mathfrak{W}^2} \times \mathfrak{W}^2 \times C}$. Therefore there is a total of $\frac{HW}{\mathfrak{W}^2}$ windows, where in each of them there is $\mathfrak{W}^2$ tokens to be fed into the self attention module, as shown in Figure \ref{fig:inter-window}. On each window-partition, self-attention is then applied by the mechanism described in Section \ref{subsub:self-attention}. Correspondingly we will have $\mathbf{Q},\mathbf{K},\mathbf{V} \in \mathbb{R}^{\frac{HW}{\mathfrak{W}^2} \times \mathfrak{W}^2 \times C}$, where each of them are the outputs of linear transformations as follows:
\begin{align}
\begin{split}
    &\mathbf{Q} = \mathbf{X}^\mathfrak{W}\mathbf{W}^Q,\\
    &\mathbf{K} = \mathbf{X}^\mathfrak{W}\mathbf{W}^K,\\
    &\mathbf{V} = \mathbf{X}^\mathfrak{W}\mathbf{W}^V,
\end{split}
\end{align}
where weights have dimensions $\mathbf{W}^Q, \mathbf{W}^K, \mathbf{W}^V \in \mathbb{R}^{C\times C}$.

Although the window-partitioning helps to reduce the computational complexity, it has deleterious effects on the long-range dependency modeling capability of the vanilla self-attention.
Here, we propose to exploit the inter-window dependencies by representing each window with a single candidate, which is the average of key and value tokens in that window. The candidates are then used to search for similarities beyond the intra-window. There will be a $\frac{HW}{\mathfrak{W}^2}$ number of candidates, defined as follows:
\begin{align}
    \mathbf{Q}^\mathfrak{W},\mathbf{K}^\mathfrak{W} \in \mathbb{R}^{\frac{HW}{\mathfrak{W}^2} \times C}.
\end{align}

After having the candidates, a resemblance matrix is utilized to measure the similarity inter-windows by comparing dot-product similarity between candidates,
\begin{align}
    \mathfrak{R}:=\mathbf{Q}^\mathfrak{W}(\mathbf{K}^\mathfrak{W})^T \in \mathbb{R}^{\frac{HW}{\mathfrak{W}^2} \times\frac{HW}{\mathfrak{W}^2}},
\end{align}
where its \emph{top-k} indices will be selected for the inter-window aggregated self-attention. The numbers tokens are kept limited by setting value of $k$ to have a reasonable amount of computations and have the computational complexity not being quadratically increasing with respect to the size of the input. The inter-window aggregated keys and values will be:
\begin{align}
\begin{split}
    &\mathbf{K}^{\mathrm{agg}}=\text{agg}[\text{top-k}(K, \mathfrak{R})] \in \mathbb{R}^{\frac{HW}{\mathfrak{W}^2} \times k\mathfrak{W}^2 \times C},\\
    &\mathbf{V}^{\mathrm{agg}}=\text{agg}[\text{top-k}(V, \mathfrak{R})] \in \mathbb{R}^{\frac{HW}{\mathfrak{W}^2} \times k\mathfrak{W}^2 \times C}.
\end{split}
\end{align}
Finally the self-attention for each window is calculated by its intact query and its aggregated top-k most similar keys and values as follows:
\begin{align}
    \text{SA}(\mathbf{X})=\text{Softmax}(\frac{\mathbf{Q}{\mathbf{K}^\mathrm{agg}}^T}{\sqrt{C}})\mathbf{V}^{\mathrm{agg}}.
\end{align}

\subsubsection{\textbf{Randomly Sampled Shift Invariant Self-Attention}}
\label{subsub:shinv}
\begin{figure}[tp]
    \centering
        \subfigure[Randomly sampled height and width shift size from categorical distribution with support of integer values less than window size, \emph{i.e.}, $1,\dots,\mathfrak{W}-1$. (\emph{multi-spectral input is shown as hyper-disks for visualization purposes only.})]{\includegraphics[width=\linewidth]{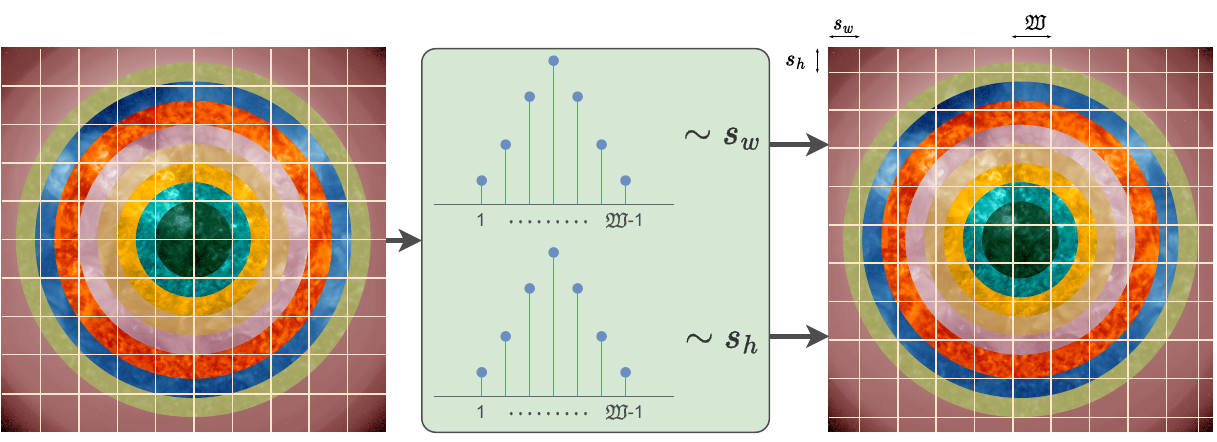}
        \label{fig:invariance:shift-sample}}
    \vfill
        \subfigure[SHift iNVariance transformer block including two consecutive transformers. First block is the vanilla window-based self-attention while the second one implements the randomly shifted window self-attention. ]{\includegraphics[width=\linewidth]{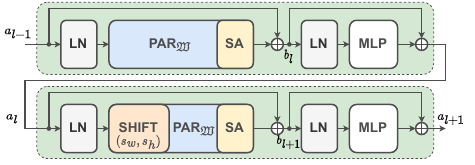}
        \label{fig:invariance:shinv}}
    \caption{Shift invaraiant transformer blocks.}
    \label{fig:invariance}
\end{figure}
When it comes to the comparison between different architectures, self-attention lacks one very important feature compared to CNNs. CNNs possess unique features such as inherent locality and weight sharing, which lead to the advantageous quality of shift invariance. These two characteristics are crucial in image compression, as pixels within nearby regions tend to display significant correlations, and achieving translation invariance is a desired trait for potential image compression techniques.


Conversely, transformer-based image compression lack compatibility with shift invariance and locality. The strategies employed by transformers to reconstruct high-resolution images can be categorized into two main patterns: the utilization of small patches with global attention \cite{chen2021ipt}, and the application of large patches with local attention \cite{wang2022uformer}. However, neither of these patterns fulfills the demands for shift invariance or locality.

The objective of the shift invariance self-attention is to enhance the transformer's ability to recognize patterns regardless of their location during translation, while also effectively utilizing local connections.  The traditional transformer structure consists of consecutive layers of self-attention (SA) and multi-layer perceptron (MLP). To enhance efficiency, the transformer typically employs local window-based attention, and a shifted window strategy is used to facilitate connections between windows. More specifically, the feature map is divided into non-overlapping windows, and self-attention is calculated within each local window, as shown in Figure \ref{fig:invariance}, and described by the following equations:
\begin{align}
\begin{split}
    b_{l} &= \text{SA}(\text{PAR}_{\mathfrak{W}}(\text{LN}(a_{l-1})))  + a_{l-1},\\
    a_l &= \text{MLP}(\text{LN}(b_l)) + b_l,\\
    b_{l+1} &= \text{SA}(\text{PAR}_{\mathfrak{W}}(\text{SHIFT}(\text{LN}(a_{l}); s_w, s_h)))  + a_{l},\\
    a_{l+1} &= \text{MLP}(\text{LN}(b_{l+1})) + b_{l+1},
\end{split}
\end{align}
where the height and width shift sizes are denoted by $s_h$ and $s_w$, respectively. In local window self-attention, the shifting sizes are fixed to half of the window size, \emph{i.e.}, $s_h=s_w=\frac{\mathfrak{W}}{2}$. This lets the consecutive transformer blocks to capture similarities over the windows \cite{liu2021swin}. The requirement to increase the number of transformer blocks to make the shifting window capture all the dependencies outside of a local window makes the vanilla shifted window attention less feasible when we expect a shift invariance functionality.
To let the window-based self-attention mechanism be performed shift-invariance we use non-deterministic shift sizes during training as shown in Figure \ref{fig:invariance:shift-sample}, where the height and width shift sizes are sampled randomly from a pre-defined categorical distribution which is peaked on the half value of window size, \emph{i.e.}, $\frac{\mathfrak{W}}{2}$. Any other distribution is possbile to be chosen, and further investigation of this could be found in Section \ref{sec:experiments:ablation}.

\section{\textbf{Experiments}}\label{sec:experiments}
\begin{figure*}[tp]
    \centering
    \subfigure{\includegraphics[width=0.45\textwidth]{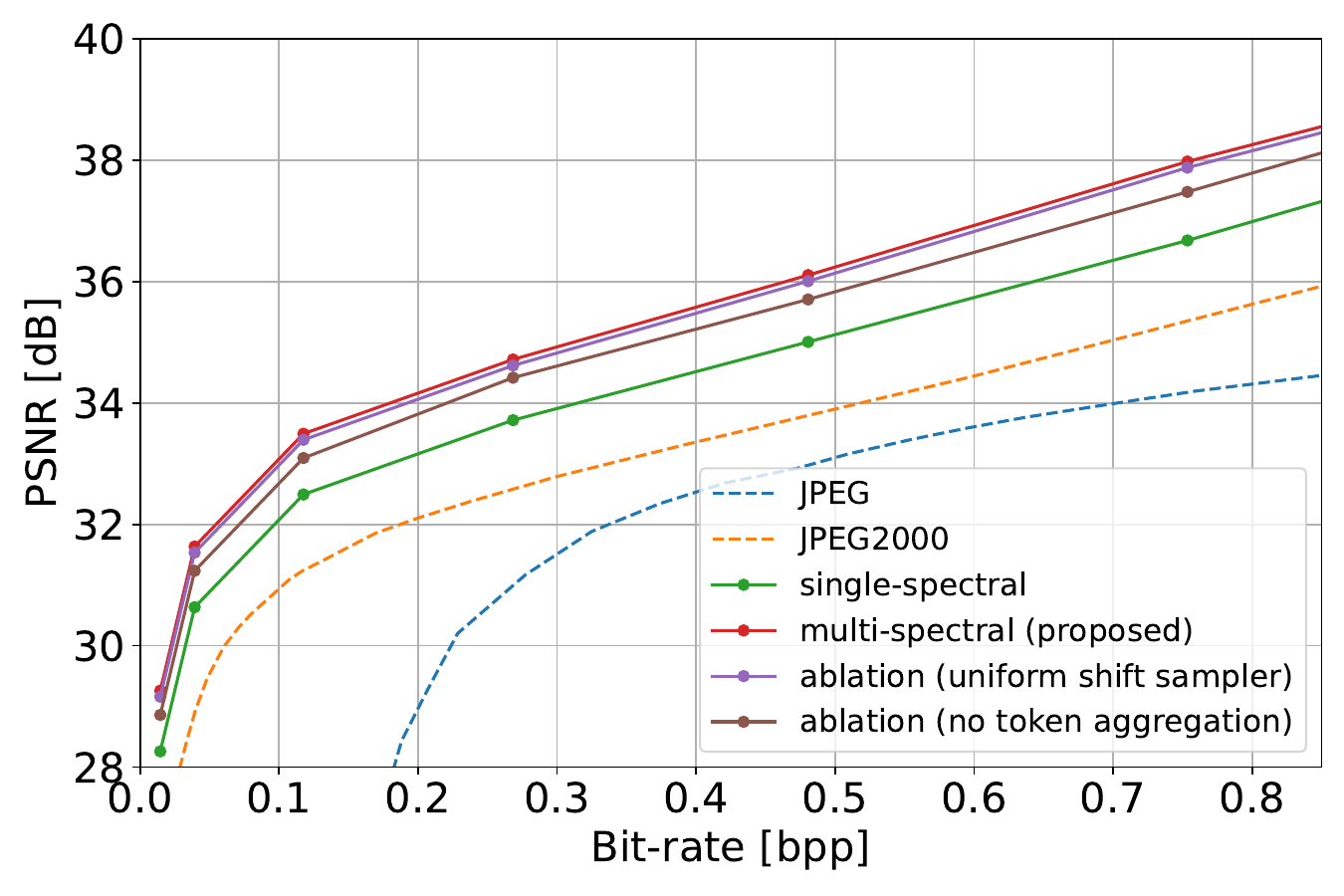}}
    \hfill
    \subfigure{\includegraphics[width=0.45\textwidth]{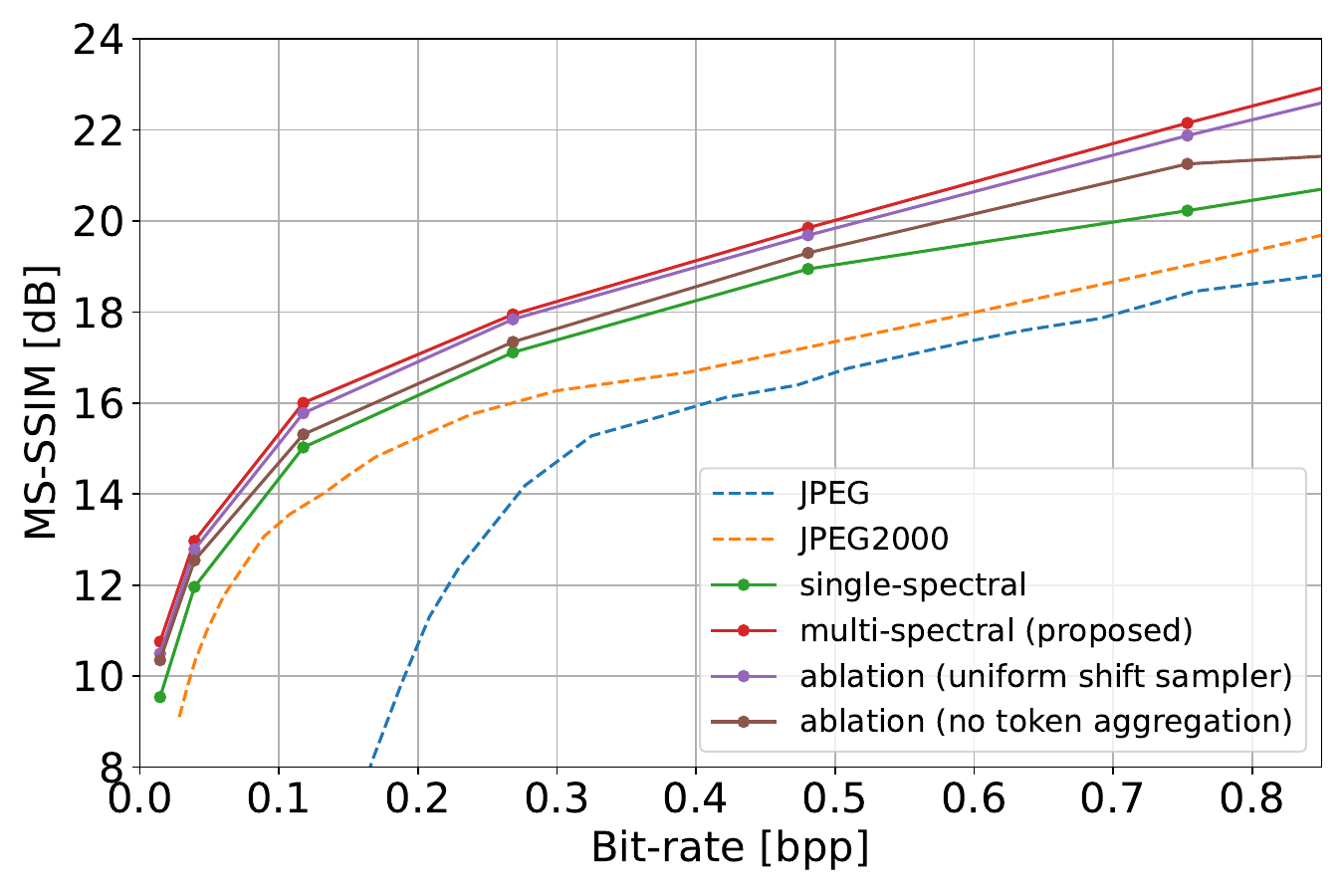}}
    \caption{The rate distortion curves aggregated over the test set are detailed in Section \ref{sec:experiments:dataset}. The left side presents PSNR (Peak Signal-to-Noise Ratio), computed from MSE (Mean Squared Error) using the formula $10\log_{10}\frac{255^2}{MSE}$. On the right side, MS-SSIM (Multi-Scale Structural Similarity Index) is displayed on a logarithmic scale using $-10\log(1-m)$ to enhance the visibility of differences. Here, $m$ represents the MS-SSIM value within the range of zero to one.}
    \label{fig:rd-cvrves}
\end{figure*}
\subsection{\textbf{Dataset}} \label{sec:experiments:dataset}
The dataset used in our experiments is based on the SDO images described in the work of Galvez et al. \cite{galvez2019}. The dataset consists of images of the Sun captured at various wavelengths, including 94, 131, 171, 193, 211, 304, 335, 1600, and 1700 \angstrom, with a temporal cadence of 6 minutes.

To reduce temporal dependencies between training samples, we downsampled the images to a cadence of 1 hour. This downsampling helps to decrease the correlation between consecutive images, allowing for more diverse training samples.
The dataset provides a comprehensive collection of solar images, capturing different aspects of the Sun's atmosphere at various wavelengths. This diversity enables us to develop and evaluate our image compression methods using a wide range of spectral information. 
To address potential biases related to solar variations at different stages of the solar cycle, we adopted a specific data division strategy inspired by the approach proposed in Salvatelli et al. \cite{salvatelli2019}. We divided the dataset based on the months in which the images were captured.

Specifically, we selected images from January to August of the years 2015 to 2018 for the training set. This range ensures coverage of different months throughout the years, helping to mitigate any biases introduced by seasonal variations in solar activity. On the other hand, we reserved images from September to December of the same years for the testing set. This separation enables us to evaluate the performance of our models on unseen data from different time periods.

In total, the training set consists of 15,768 images where each image contains 9 wavelengths at its channel dimension, while the test set comprises 3,315 images. The results reported in this section are based on the evaluation of our models on the test set, providing insights into their performance on previously unseen data.

\subsection{\textbf{Implementation Specifications}} \label{sec:experiments:implementation}
We trained a total of seven models, each with a different hyper-parameter $\lambda$ governing the rate-distortion trade-off as defined in Equation (\ref{eq:rate-distortion}). The chosen values for $\lambda$ were empirically determined as $\{0.0015, 0.0035, 0.0070, 0.0125, 0.0250, 0.0410, 0.0550\}$, and each model was trained for 200 epochs.

For training, we used the Adam optimizer \cite{kingma15adam} with a batch size of 8. The training data consisted of randomly cropped patches of size $256\times256$ from the original $512\times512$ images. The initial learning rate was set to $10^{-4}$ and annealed during training to $1\times10^{-5}$ to facilitate convergence.


During the evaluation phase, we performed entropy coding of the latent integer values using range asymmetric numeral systems coding \cite{duda2013asymmetric}. It's important to note that this entropy coding is lossless and does not impact the measured performance or functionality of the algorithm during training. It is only during the evaluation phase that entropy coding is applied, as it allows for comparison with standard codecs such as JPEG \cite{wallace1992jpeg} and JPEG-2000 \cite{jpeg2000}.

\subsection{\textbf{Ablation Study}} \label{sec:experiments:ablation}
To verify the impact of both aggregating tokens inter-windows and non-deterministic shift sizes, we have conducted two isolated ablation studies. In the first set of models, the token-aggregation transformer blocks are replaced with vanilla Swin transformer blocks \cite{liu2021swin} to see how aggregation of windows could play more effective role than the local window-limited attentions even when the shifts are in place. The second ablative study is concerned with how the distribution of the shift sizes could affect the rate-distortion perfomance of the multi-spectral compressor. We have used a uniform distribution with support set of $\{1,\dots,\mathfrak{W}-1\}$ instead of the pre-defined categorical distribution with the same support set, in the original network. As is shown in Figure \ref{fig:rd-cvrves}, the choice of seeing outside of the local window (by aggregating tokens) has much tangible influence than how we choose to sample the shift sizes of the windows.

\section{\textbf{Conclusion}}\label{sec:conclusion}
In this work a multi-spectral transformer-based neural image compression algorithm were proposed to effectively improve the rate-distortion performance and save the communication cost in data-intensive missions studying solar dynamics. Downloaded data from these missions possess high redundancy over the frequency spectrum and we proposed mechanisms to remove redundancies effectively. Inter-window token aggregation and shift invariance window partitioning were added to the transformer blocks to better de-correlate the high dimensional multi-spectral images of the Sun.

\section*{\textbf{Acknowledgment}}
This research is based upon work supported by the National Aeronautics and Space Administration (NASA), via award number 80NSSC21M0322 under the title of \emph{Adaptive and Scalable Data Compression for Deep Space Data Transfer Applications using Deep Learning}.
\bibliographystyle{IEEEtran}
\bibliography{mybib}
\end{document}